\documentclass[sigconf]{acmart}

\usepackage{graphicx}
\usepackage{subfigure}
\usepackage{url}
\usepackage{amsmath}
\AtBeginDocument{%
  \providecommand\BibTeX{{%
    \normalfont B\kern-0.5em{\scshape i\kern-0.25em b}\kern-0.8em\TeX}}}

\setcopyright{acmcopyright}
\copyrightyear{2019}
\acmYear{2019}
\acmConference[WI '19 Companion]{IEEE/WIC/ACM International Conference on Web Intelligence}{October 14--17, 2019}{Thessaloniki, Greece}
\acmBooktitle{IEEE/WIC/ACM International Conference on Web Intelligence (WI '19 Companion), October 14--17, 2019, Thessaloniki, Greece}
\acmPrice{15.00}
\acmDOI{10.1145/3358695.3360930}
\acmISBN{978-1-4503-6988-6/19/10}

\begin{document}

\title{User's Centrality Analysis for Home Location Estimation}

\author{Shiori Hironaka}
\email{s143369@edu.tut.ac.jp}
\orcid{0000-0001-7994-2858}
\affiliation{%
  \institution{Toyohashi University of Technology}
  \streetaddress{1--1 Hibarigaoka, Tempaku-cho}
  \city{Toyohashi}
  \state{Aichi}
  \country{Japan}
  \postcode{441--8580}
}

\author{Mitsuo Yoshida}
\email{yoshida@cs.tut.ac.jp}
\orcid{0000-0002-0735-1116}
\affiliation{%
  \institution{Toyohashi University of Technology}
  \streetaddress{1--1 Hibarigaoka, Tempaku-cho}
  \city{Toyohashi}
  \state{Aichi}
  \country{Japan}
  \postcode{441--8580}
}

\author{Kyoji Umemura}
\email{umemura@tut.jp}
\affiliation{%
  \institution{Toyohashi University of Technology}
  \streetaddress{1--1 Hibarigaoka, Tempaku-cho}
  \city{Toyohashi}
  \state{Aichi}
  \country{Japan}
  \postcode{441--8580}
}

\renewcommand{\shortauthors}{Hironaka et al.}

\begin{abstract}
  User attributes, such as home location, are useful for many applications.
  Many researchers have been tackling how to estimate users' home locations using relationships among users.
  It is known that the home locations of certain users, such as celebrities, are hard to estimate using relationships.
  However, because estimating the home locations of all celebrities is not actually hard, it is important to clarify the characteristics of users whose home locations are hard to estimate.
  We analyze whether centralities, which represent users' characteristics, and the tendency to have the same home locations as friends are related.
  The results indicate that PageRank and HITS scores are related to whether users have the same home location as friends, and that users with higher HITS scores have the same home location as their friends less often.
  This result indicates that there are two types of users whose home locations are difficult to estimate: hub users who follow many celebrities and authority users who are celebrities.
\end{abstract}



\maketitle

\section{Introduction}

Social media is widely used to interact with friends, get information, and post news.
Thus, social media data represent preferences and social trends, and these data are used for various applications.
User attributes relating to the real world, such as the home locations of users, are necessary in many applications, such as trend detection for marketing~\cite{Benhardus2013}, news recommendations for user experience~\cite{Jonnalagedda2013,Phelan2009}, quantitative observations of how disease is spreading in the real world~\cite{Signorini2011}, and real-world event detection~\cite{Sakaki2010}.
Although users' real attributes are important, they are not provided in many cases.
Among real attributes, home location is widely used to ground user information in computational social science studies.

A user's home location is often obtained from geo-tagged tweets posted by the user.
However, it has been reported that only 3.1\% of users post geo-tagged tweets on Twitter~\cite{Sloan2015}.
Among Japanese-language users, only 0.8\% post geo-tagged tweets.
It is necessary to estimate unknown users' home locations from the known home locations of other users.

A social graph representing relationships between users is used to estimate a user's home location~\cite{Jurgens2015}.
The assumption is that the geographic distances between connected users on the social graph are close.
However, Rahimi et al.~\cite{Rahimi2015} showed that the home locations of some users, such as celebrities mentioned by many users, are hard to estimate.
Ebrahimi et al.~\cite{Ebrahimi2018} showed that not all celebrities' home locations are hard to estimate---only those of global celebrities, who are mentioned geographically widely.
We consider that users who are close on the social graph but not geographically close share certain features.
Many centrality measures evaluate or compare the importance of the nodes on a graph.
It is considered that centrality scores, such as degree centrality, PageRank~\cite{Page1999}, and HITS scores~\cite{Kleinberg1999}, are related to the tendency for neighbors to have similar attributes as their neighbors on the graph.

In this paper, we analyze whether various centrality scores and the tendency to have the same home location as friends are related.
We found that users who have the same home location as many friends have different PageRank and HITS scores.
This result indicates that there are two types of users whose home locations are difficult to estimate: hub users who follow many celebrities and authority users who are celebrities.

\section{Data}

In this section, we describe the dataset for analysis.
The dataset consists of home location data and social graph data.

\subsection{Home Location}

We assume that users mainly post tweets around their home location; thus, we define the home location of a user as the city from which the user posts tweets most frequently.
We collected geo-tagged tweets posted within the rectangle covering Japan\footnote{The area within latitude 20 to 50 degrees north and longitude 110 to 160 degrees east.} in 2014 using Twitter Streaming API with location parameters.
We then identified the city (the Japanese municipality) containing the coordinates of the collected geo-tagged tweets using the boundary data provided by the Statistics Bureau of Japan's Ministry of Internal Affairs and Communications.\footnote{\url{https://www.e-stat.go.jp/en} (viewed 2019-07-19)}

We assigned the users' home location as the location from which they most frequently posted,
and for the sake of accuracy, we included
only users who posted geo-tagged tweets more than five times in the same area.
Consequently, we obtained the home locations of 471,761 users (contains 1,873 unique cities).

\subsection{Social Graph\label{sec:socialgraph}}

To construct a social graph, we use the \textit{following} relationships of users who were assigned a home location.
We collected their followees and followers in July 2015.
We construct a social graph by creating a directed edge from user X to user Y when user X is following user Y.
We removed the edges to and from the users who are not assigned a home location.
The constructed social graph is a simple directed graph.
As a result, the social graph contained 471,761 nodes and 8,295,355 edges.
All nodes in this graph have known home locations.
The average number of incoming and outgoing edges was 17.58.
The average number of mutual followers was 13.2,
and 42,316 users had no incoming and no outgoing edges.

\section{Analysis Method}

We analyze relationships between the centralities of users and the tendency to have the same home locations as their friends.
In this section, we describe how to calculate the centrality scores and how to measure the tendency.
We then calculate the degree of bias, to clarify the difference in the distributions of centrality scores by the tendency.

\subsection{Centrality Measure}

We use the in-/out-degree centrality, PageRank~\cite{Page1999}, and the authority and hub of the HITS algorithm~\cite{Kleinberg1999} for analysis.
The in-degree centrality has a larger score when the user has more followers, and the out-degree centrality has a larger score when the user has more followees.
The degree is used as a measure of celebrity in previous studies~\cite{Rahimi2015,Ebrahimi2018}.
Similar to in-degree centrality, the PageRank score increases as a user is followed by users with higher scores.
We expect that not only users who have many followers but also users who are being followed by such users will have higher scores.
The HITS algorithm assumes that a hub user follows many authority users, and that an authority user is being followed by many hub users; then, the scores are calculated.
The hub score tends to be high if the user is following many users, and the authority score tends to be high if the user is being followed by many users.
PageRank has been used to identify the influential users~\cite{Kwak2010}.
We consider that HITS is as effective as PageRank.
We calculate the centralities on the social graph constructed in section~\ref{sec:socialgraph} using NetworkX.\footnote{\url{https://networkx.github.io/} (viewed 2019-04-15)}

\subsection{User Groups Based on Neighborhood Friends}

We use the location estimation method proposed by Davis Jr. et al.~\cite{DavisJr2011} as an 
indicator of the proximity between users' home locations and those of their friends.
The method estimates the user's location as the most frequent location among the home locations of their friends.
Using the estimation results, we categorize users into the following three groups, which represent the strength of their tendency to have the same home location as their friends:
(a)~users whose home location is estimated correctly, (b)~users whose home location is estimated incorrectly, and (c)~users whose home location cannot be estimated.
It can be assumed that these users are, respectively, (a)~users labeled as \textit{easy}, who have the same home location as the majority of their friends, (b)~users labeled as \textit{hard}, who do not have the same home location as the majority of their friends, and (c)~users labeled as \textit{unknown}, whose data offer no clues concerning their home location, and whose proximity to their friends cannot be measured.

In this paper, we define friends as users who are mutual followers.
We use the estimation result by leave-one-out cross-validation.

\subsection{Calculation of Distribution}

The distribution of the centrality score for user set $U$ is calculated as follows.
We define the total number of users as $|U| = N$, and the number of users within an interval $i: [x_i, x_{i+1})$ is $n_i$.
Then, the percentage that the user exists in interval $i$ is $n_i / N$.
We call $f(i) = n_i / N$ a score distribution.
The intervals are chosen to satisfy $\sum_i n_i / N = 1$.

We calculate the score distributions of all users and of each group.
Then, to clarify the degree of bias of the user group divided by the tendency,
we calculate the difference in the distributions between each user group and all users with the following method.
Assuming that the percentage of all users in interval $i$ is $a$, and that the percentage of users to be compared in interval $i$ is $b$, the degree of bias of the distribution of all users is calculated as $\log_{10}(b/a)$.
It becomes negative when the percentage of users in the distribution of all users is higher, and positive when the percentage is lower.
The degree of bias becomes large if the difference between the compared distributions is large.

\section{Results and Discussion}

\begin{figure*}[tp]
  \centering
  \subfigure[Distribution (in-degree centrality)]{\includegraphics[width=.49\linewidth]{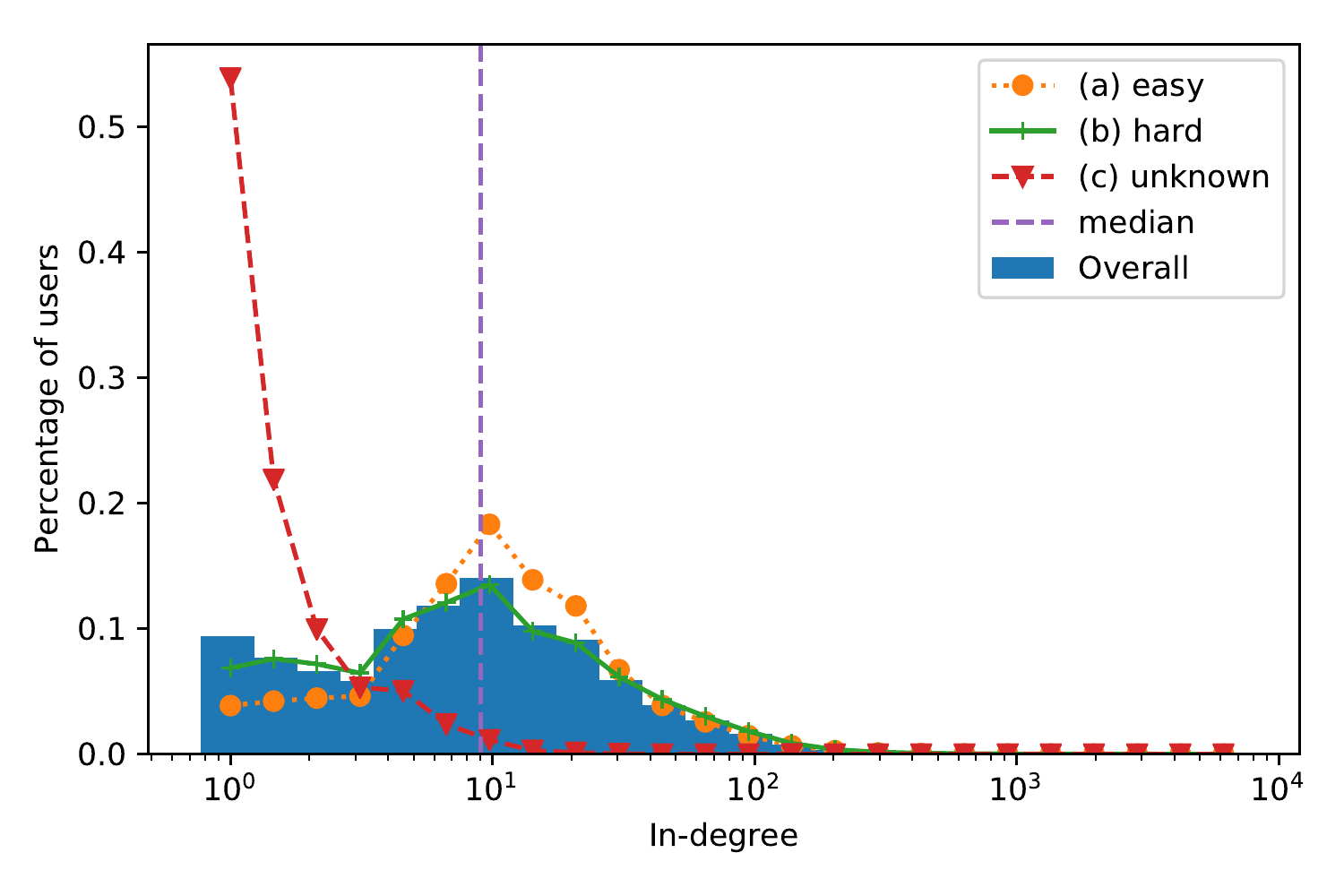}}%
  \subfigure[Difference (in-degree centrality)]{\includegraphics[width=.49\linewidth]{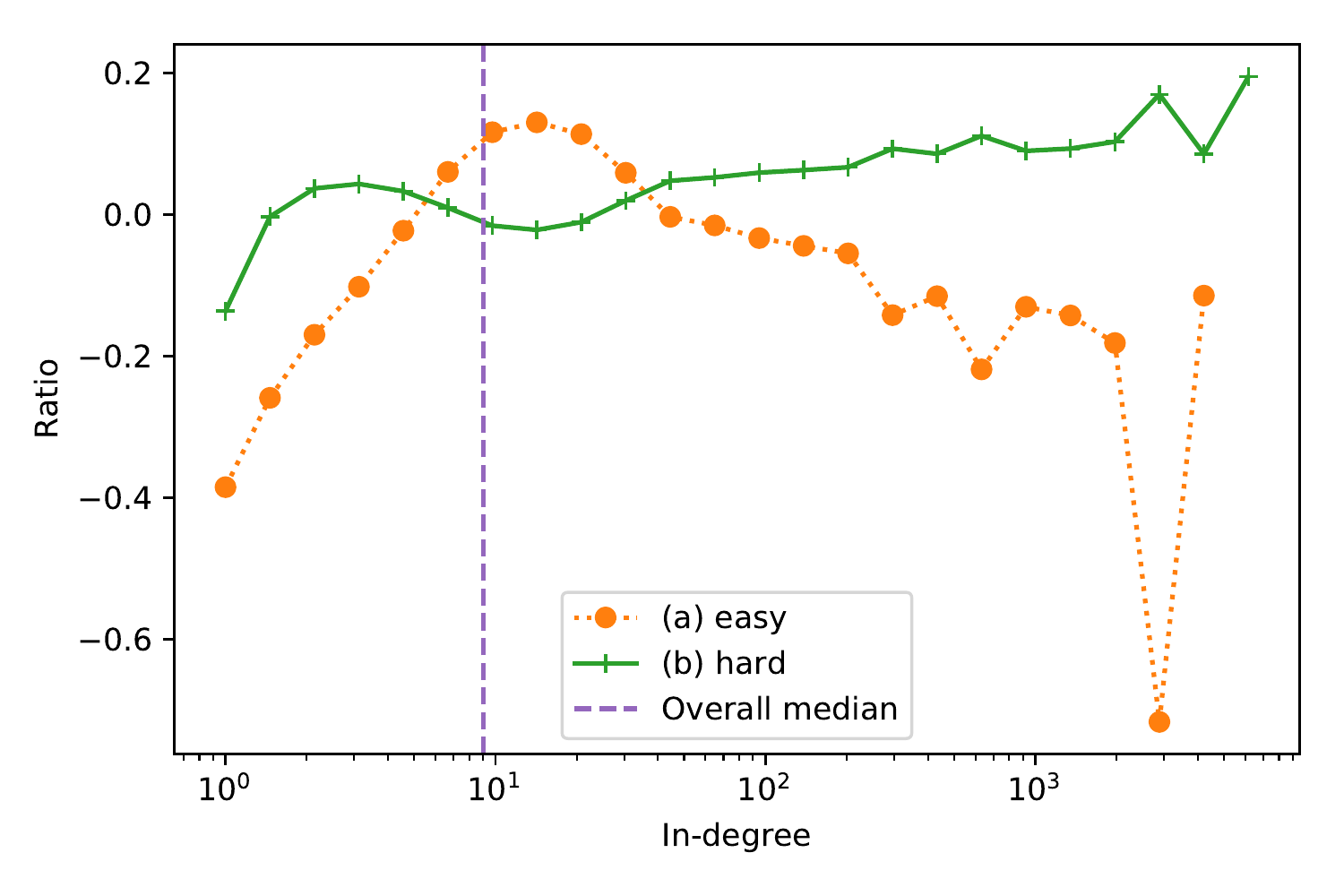}}
  \subfigure[Distribution (out-degree centrality)]{\includegraphics[width=.49\linewidth]{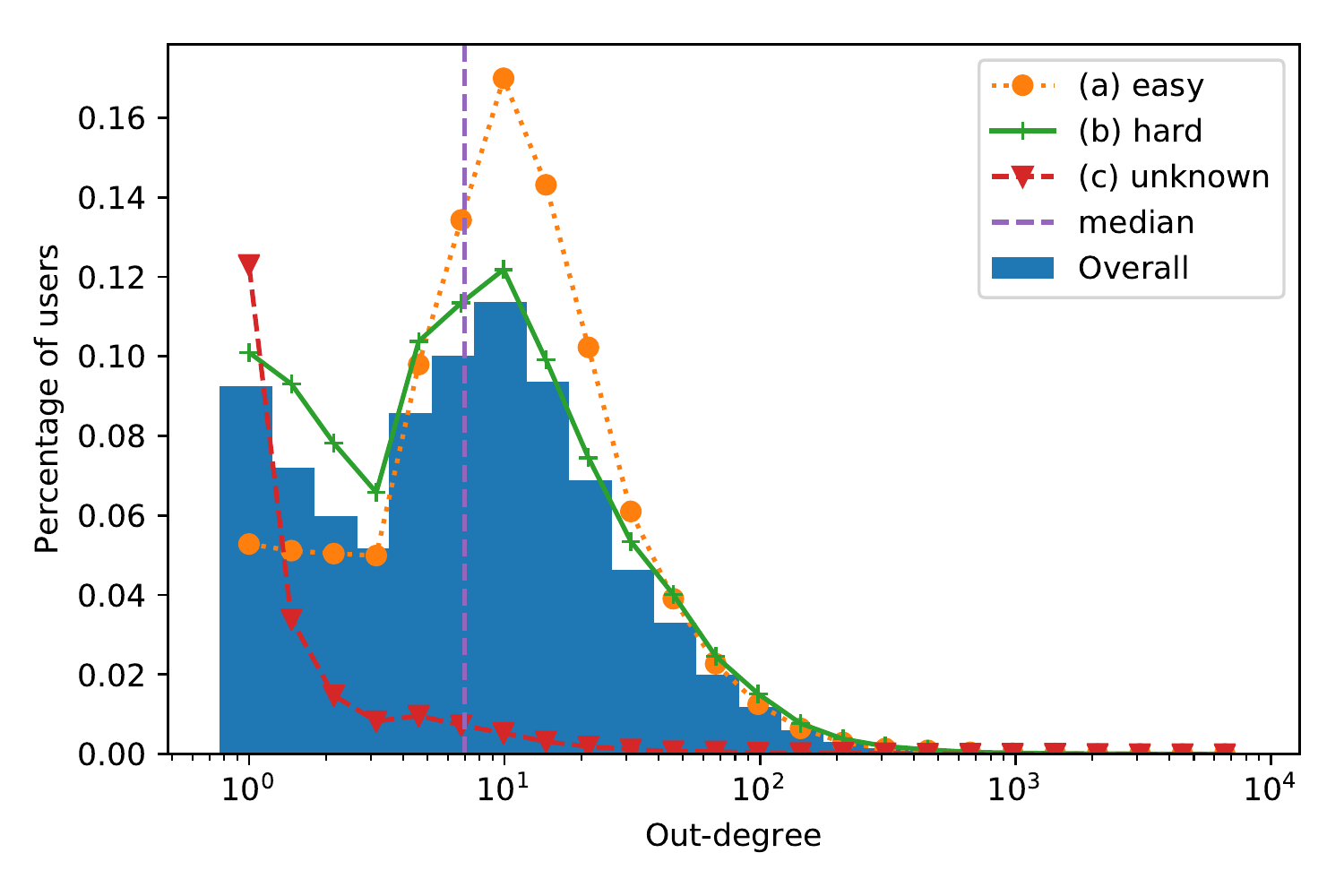}}%
  \subfigure[Difference (out-degree centrality)]{\includegraphics[width=.49\linewidth]{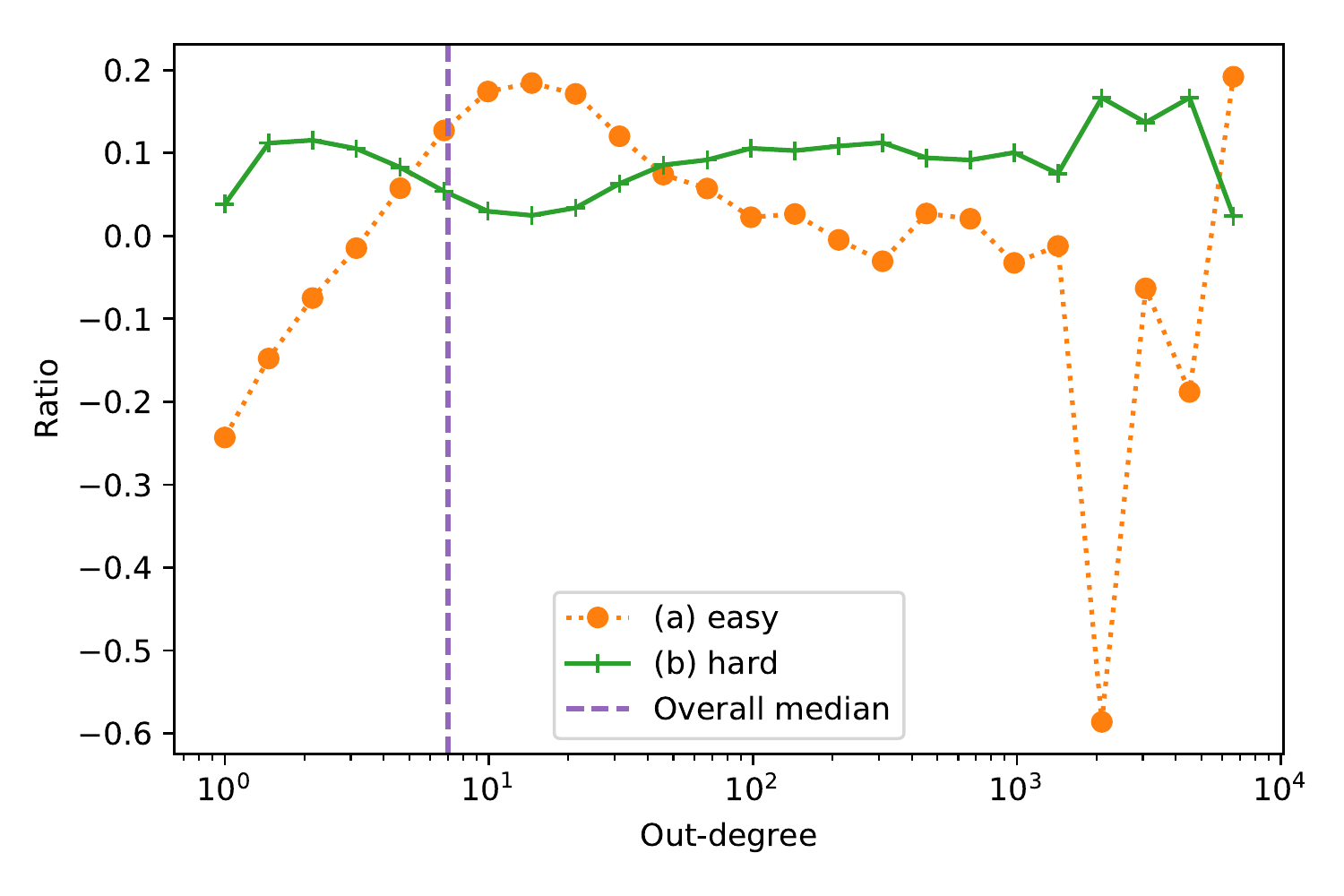}}
  \Description{description}
  \caption{Distributions and differences of in-/out-degree centrality scores. The peak positions of the distributions of (a) and (b) are not very different.}
  \label{fig:degree}
\end{figure*}

\begin{figure*}[p]
  \centering
  \subfigure[Distribution (PageRank)]{\includegraphics[width=.49\linewidth]{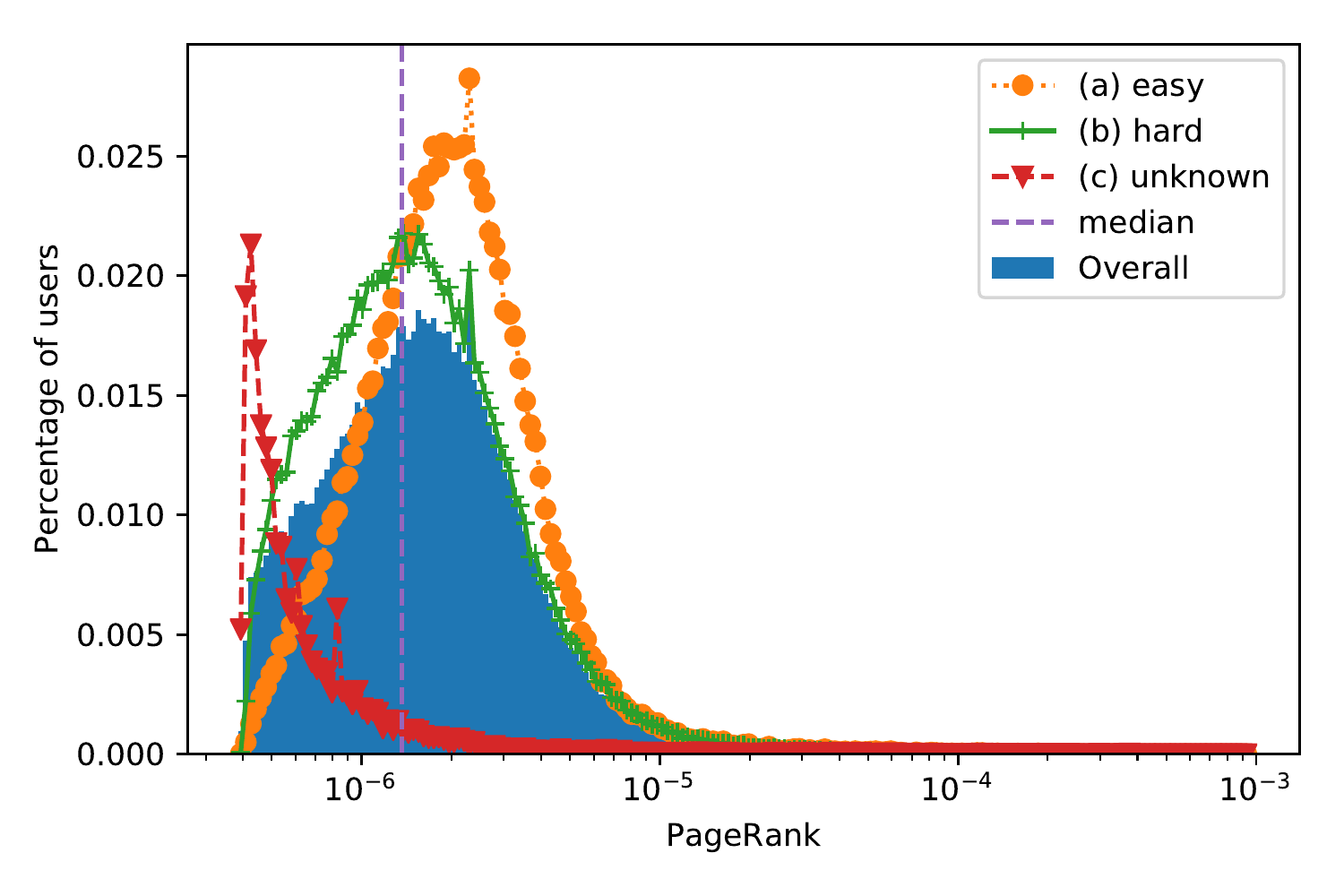}}%
  \subfigure[Difference (PageRank)]{\includegraphics[width=.49\linewidth]{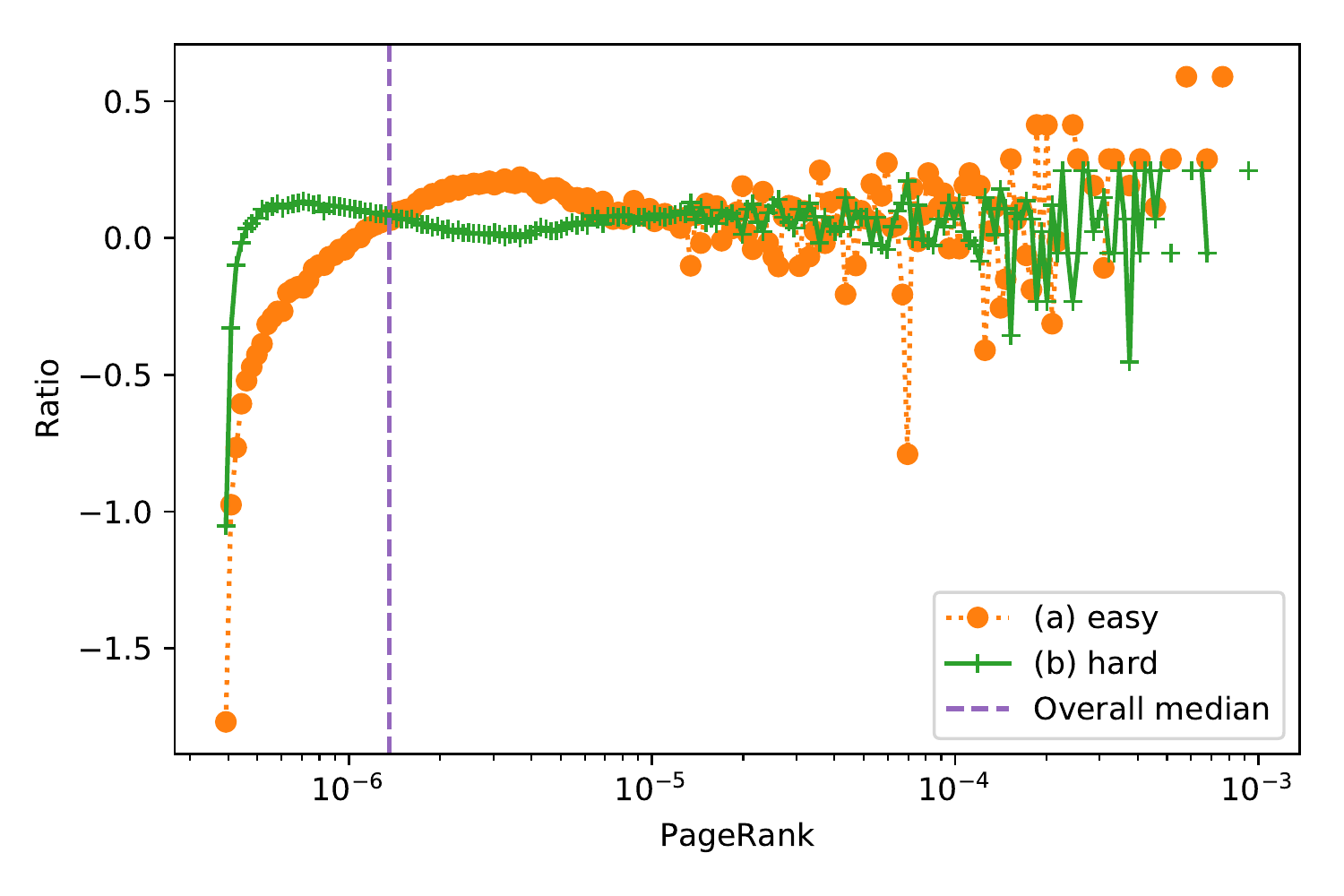}}
  \subfigure[Distribution (authority of HITS)]{\includegraphics[width=.49\linewidth]{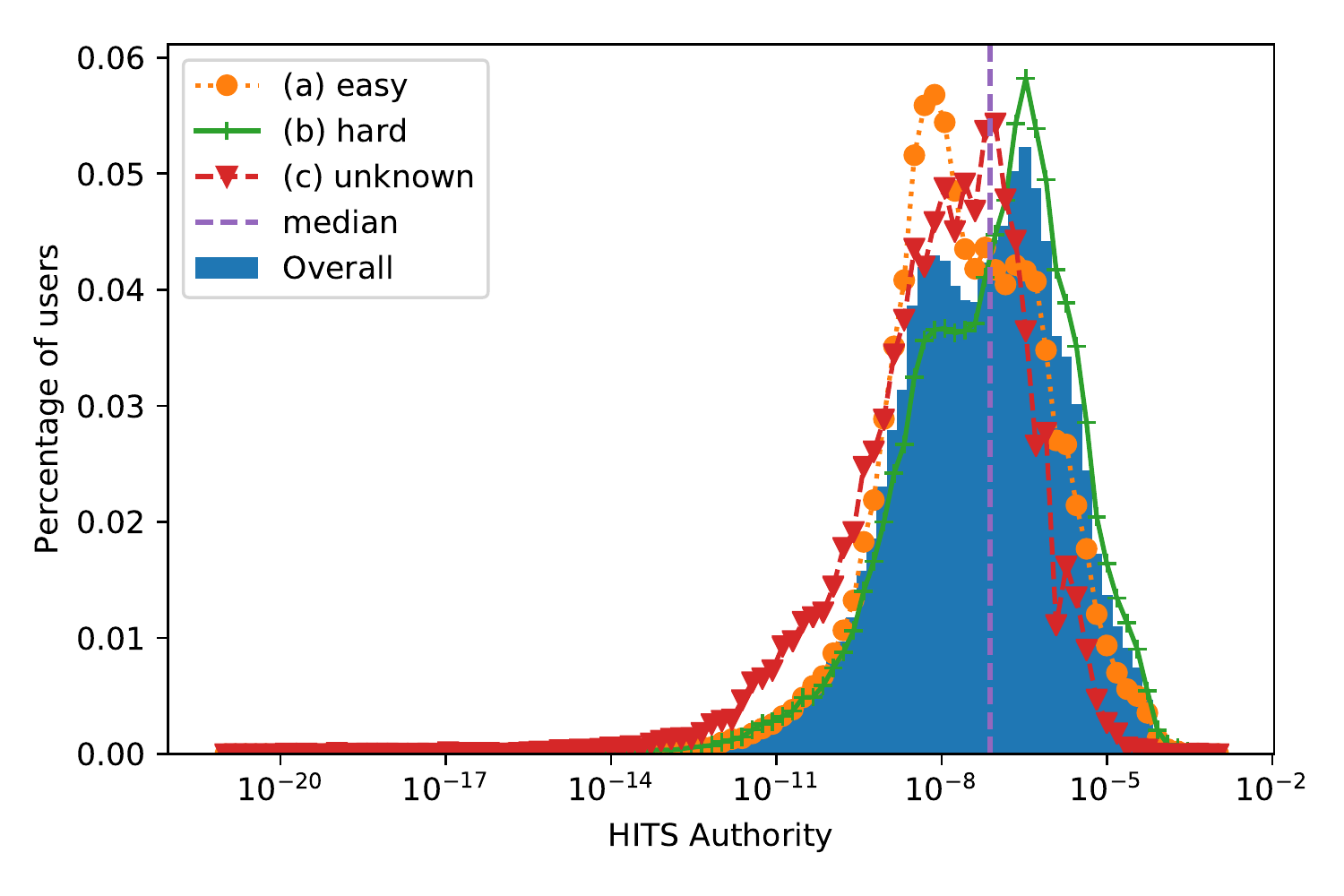}}%
  \subfigure[Difference (authority of HITS)]{\includegraphics[width=.49\linewidth]{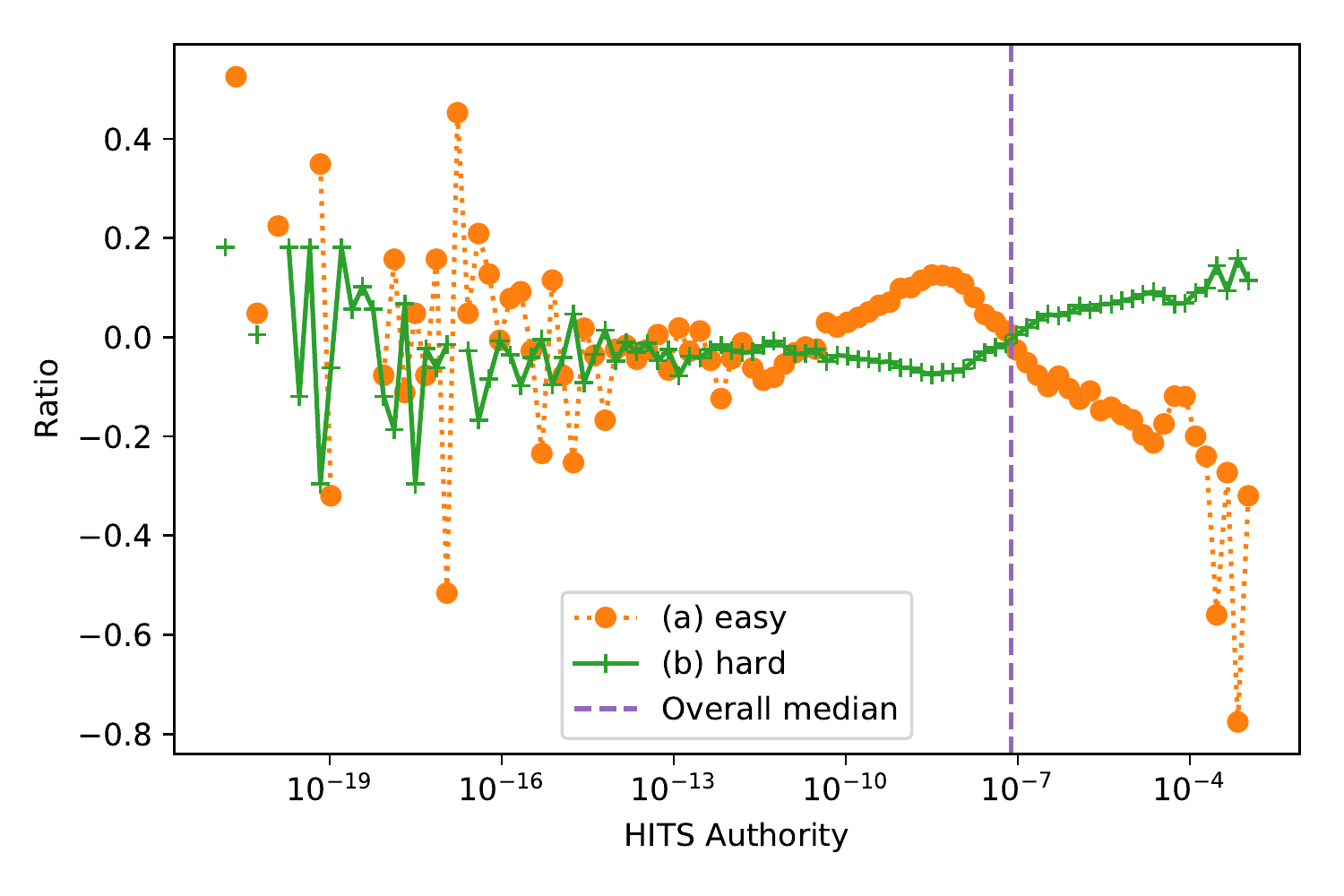}}
  \subfigure[Distribution (hub of HITS)]{\includegraphics[width=.49\linewidth]{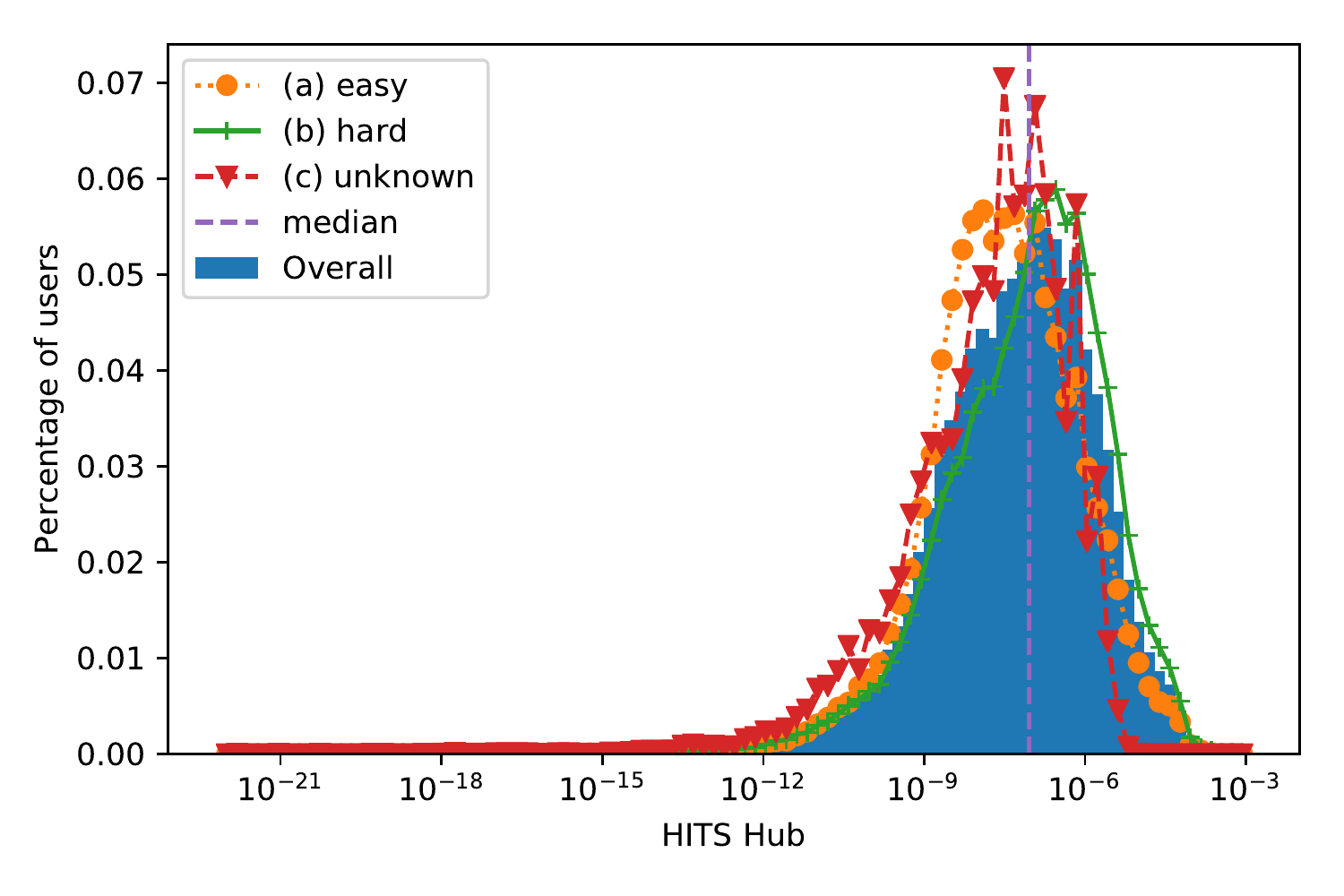}}%
  \subfigure[Difference (hub of HITS)]{\includegraphics[width=.49\linewidth]{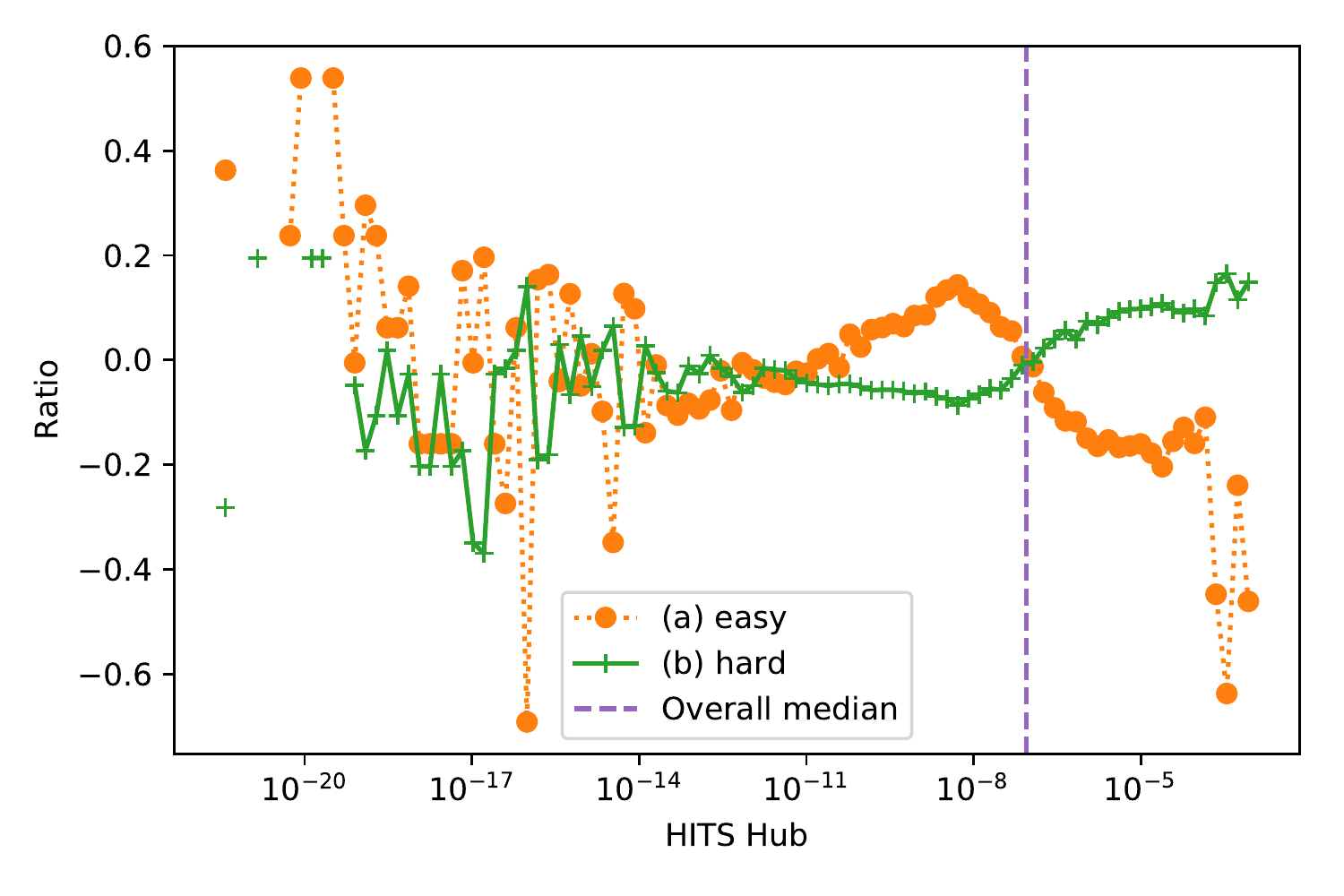}}
  \Description{description}
  \caption{Distributions and differences of PageRank and HITS scores. The users in (a) tend to have higher PageRank scores than the users in (b). For the HITS scores, the distribution of (a) is located on the right side (indicating a large value), and the distribution of (b) on the left side (indicating a small value), with a median value as the center. The minimum score of PageRank is outside the graph.}
  \label{fig:pagerank-hits}
\end{figure*}

We categorized all users (471,761 users) who were assigned home locations into three groups, as follows:
(a) 121,275 users had the same home location as the majority of their friends, (b) 267,809 users had a different home location from the majority of their friends, and (c) 82,677 users had no friends with a home location.
We calculated the score distributions of all users (Overall) and the three groups for each centrality.
We then calculated the degree of bias of (a) and (b) for each centrality.\footnote{We excluded (c) users because their tendencies cannot be measured.}
Figs.~\ref{fig:degree} and \ref{fig:pagerank-hits} show the distribution of the in- and out-degree centrality, PageRank, and the authority and hub of the HITS scores.

As the results of the in- and out-degree centralities illustrated in Fig.~\ref{fig:degree} show the peak positions of the distributions of (a) and (b) are not very different.
In terms of the ratio to the overall distribution, the figure shows that the ratio of (a) increases when the value is around 20.
Subsequently, as the centrality score becomes larger, the ratio of (a) decreases.
If users with many followees or followers are considered celebrities, it is difficult to estimate the home locations of the celebrities correctly.
Davis Jr. et al.~\cite{DavisJr2011} reported that the precision is highest when estimated using only users with a number of mutual followers between 20 and 200.
However, we cannot compare the results, because the methods of counting the number of mutual followers are different.

Fig.~\ref{fig:pagerank-hits} illustrates the results for PageRank and HITS.
As the figure shows, the users in (a) tend to have higher PageRank scores than the users in (b).
The authority and hub HITS scores seem to exhibit a difference in the peak position of distribution between (a) and (b).
In both, the peak position of (a) is located on the right side (indicating a large value), and the peak position of (b) on the left side (indicating a small value), with a median value as the center.
Because the HITS scores have a smaller distribution overlap between (a) and (b) than do PageRank scores,
whether Twitter users have same home locations as their friends can be better understood by looking at the HITS scores.
This result suggests that the social graph of Twitter has two types of users: hub users who follow many celebrities and authority users who are celebrities.
In the results for the authority and hub HITS scores, there are few users in (a) with high scores.
These results indicate that users who have either high authority or high hub HITS scores have low proximity with the home locations of friends, not only users who have both high authority and high hub HITS scores.

\section{Conclusion}

We analyzed the relationships between the tendency in home locations between users and their friends, and their centrality scores, using in-/out-degree centralities, PageRank, and the HITS algorithm.
The results show that users who have the same home location as a small number of friends have a high PageRank score.
Because users who have high HITS scores have the same home location as only a small number of friends, there are two types of users whose home locations are difficult to estimate: hub users who follow many celebrities and authority users who are celebrities.
We found that users who have either high authority or high hub HITS scores have low proximity to their friends' home locations, not only users who have both high authority and high hub HITS scores.

\bibliographystyle{ACM-Reference-Format}
\bibliography{./references}

\end{document}